\documentclass[twocolumn,preprintnumbers,amsmath,amssymb]{revtex4}
\input{epsf}

\usepackage{natbib}
\usepackage[dvips]{graphicx}
\usepackage{mciteplus}

\usepackage{subfigure}
\usepackage{dcolumn}
\usepackage{bm}
\usepackage{latexsym}

\newcommand{\gsim}{\gtrsim}
\newcommand{\lsim}{\lesssim}
\newcommand{\ie}{\textsl{i.e.~}}

\def\spose#1{\hbox to 0pt{#1\hss}}
\def\lta{\mathrel{\spose{\lower 3pt\hbox{$\mathchar"218$}}
     \raise 2.0pt\hbox{$\mathchar"13C$}}}
\def\gta{\mathrel{\spose{\lower 3pt\hbox{$\mathchar"218$}}
     \raise 2.0pt\hbox{$\mathchar"13E$}}}

\newcommand{\mpl}{m_\mathrm{Pl}}
\newcommand{\de}[2]{\kern - #1 em \mathrm{d} #2}

\newcommand{\simle}{\,{}^<_{\sim}\,}

\begin{document}

\title{Collapse of Small-Scale Density Perturbations during Preheating
  in Single Field Inflation}

\author{Karsten Jedamzik} \email{jedamzik@lpta.univ-montp2.fr}
\affiliation{Laboratoire de Physique Th\'eorique et Astroparticules, 
UMR 5207-CNRS, Universit\'e de Montpellier II, F-34095 Montpellier, France}

\author{Martin Lemoine} \email{lemoine@iap.fr} \affiliation{Institut
d'Astrophysique de Paris, \\ UMR 7095-CNRS, Universit\'e Pierre et Marie
Curie, \\ 98bis boulevard Arago, 75014 Paris, France}

\author{J\'er\^ome Martin} \email{jmartin@iap.fr}
\affiliation{Institut d'Astrophysique de Paris, \\ UMR 7095-CNRS,
Universit\'e Pierre et Marie Curie, \\ 98bis boulevard Arago, 75014
Paris, France}

\date{\today}

\begin{abstract}
  After cosmic inflation and before the transition to radiation
  domination, the cosmic energy density may have been dominated during
  an extended period by an oscillating massive scalar condensate. We
  show that during this period, sub-Hubble scale perturbations are
  subject to a metric preheating instability in the narrow resonance
  regime. We analyze in detail both, quadratic and quartic potentials.
  The instability leads to the growth of density perturbations which
  in many cases become non-linear already before the beginning of a
  radiation dominated Universe.  This is particularly the case when
  requiring a phenomenologically preferred low reheat
  temperature. These early structures may lead to the emission of
  gravitational waves and the production of primordial black
  holes. Furthermore, it is not clear if they could modify the
  prediction of linear curvature perturbations on very large scales.
\end{abstract}

\maketitle

\section{Introduction}

The Universe having passed through an early inflationary epoch is by
now a widely accepted paradigm for an explanation of apparent cosmic
large-scale homogeneity as well as the perturbations observed in the
cosmic microwave background radiation (CMBR). One of the earliest
incarnations of inflation was given in Ref.~\cite{Linde:1983gd} where
a massive scalar field with potential $V(\phi) = m^2\phi^2/2$,
starting from super-Planck values, slowly rolls down towards its
minimum thereby driving a period of accelerated expansion. Though many
alternate scenarios of inflation have been discussed over the years,
the above scenario of ``chaotic inflation'' remains one of the most
successful in terms of reproducing the observed
perturbations~\cite{Linde:2007fr}. In particular, the predicted
spectral index $n_{_{\rm S}}\simeq 0.96$ in this model is consistent
with that inferred from the WMAP-7 analysis of the CMBR
inhomogeneities, which gives $n_{_{\rm S}}\simeq 0.963\pm 0.012$
($68\%$ CL)~\cite{Komatsu:2010fb}.

\par

After inflation has ended, cosmic energy density is dominated by an
almost homogeneous oscillating inflaton field. A transition towards a
radiation dominated Universe, as required to exist at the latest
shortly before the epoch of Big Bang nucleosynthesis (BBN) which
starts at cosmic temperature $T\sim 1\,$MeV, is usually assumed to
occur by the perturbative decay of massive inflaton particles into
radiation. This epoch is referred to as the epoch of reheating.
Parametric resonance phenomena, inducing energy transfer to other
scalars coupled to the inflaton, referred to as preheating, though
rich in their physics, have been shown to fail to provide a complete
transition towards a radiation dominated Universe (see,
e.g.~\cite{Kofman:1997yn,Bassett:2005xm} for a review of preheating
models). In case the inflaton would decay immediately after the end of
chaotic inflation, and assuming $m^2\phi^2/2$ inflation, the cosmic
reheat temperature would be exceedingly large, \ie $T_{\rm rh}\sim
10^{16}$GeV. Such a large $T_{\rm rh}$ is phenomenologically
problematic, as likely leading to the re-generation of unwanted
relics, such as magnetic monopoles at a GUT transition, or massive
gravitinos in
supergravity~\cite{Ellis:1982yb,Nanopoulos:1983up}. Modern constraints
from BBN on unstable gravitinos constrain this reheating temperature
down to $T_{\rm rh}\lsim 10^{7}\,$GeV~\cite{Jedamzik:2006xz,
  Kawasaki:2008qe,Bailly:2009pe}. It is thus possible (and desirable)
that between the end of chaotic inflation and reheating cosmic energy
density is dominated for a very long expansion period by an almost
homogeneous inflaton condensate. For example, for $T_{\rm rh}$ as low
as $\sim 1\,$MeV the Universe would have expanded an immense factor
$\sim 10^{25}$ between the end of inflation and reheating.  It is such
a cosmic era which is the topic of the current paper.

\par

An inflaton oscillatory phase after inflation has received some
attention with respect to the evolution of super-Hubble size
perturbations~\cite{Nambu:1996gf,Hamazaki:1996ir,Finelli:1998bu}.  It
has been shown that, though one finds parametric resonance in the
$m^2\phi^2/2$ case, super-Hubble size perturbations thought to be
responsible for the anisotropies in the CMBR stay constant in
amplitude, as is the case in a radiation dominated Universe. It was
therefore concluded that in single-field models ``metric preheating''
does not lead to non-trivial changes in the evolution of
perturbations. However, as we will see in this paper (see also
Ref.~\cite{Nambu:1996gf}), parametric resonance occurs even for
sub-Hubble size regions, \ie for all modes with physical wave vectors
$k_{\rm phys}<\sqrt{3 H m}$ (for a quadratic potential), where $H$ is
the Hubble constant. In fact, these modes lie in the narrow resonance
regime and the red-shifting is such that any mode which had once been
resonant will always remain resonant. Due to this resonance the
curvature perturbations of sub-Hubble modes stay constant rather than
decreasing. One can readily show that this implies that the density
contrast grows with scale factor $\delta \rho _{\bm k}/\rho\propto a$
implying non-linear cosmic structure formation after an expansion
factor of $\sim 10^6$. This fact has so far received little attention
in the literature.

\par

\begin{figure}
  \centering
  \includegraphics[width=0.5\textwidth,clip=true]{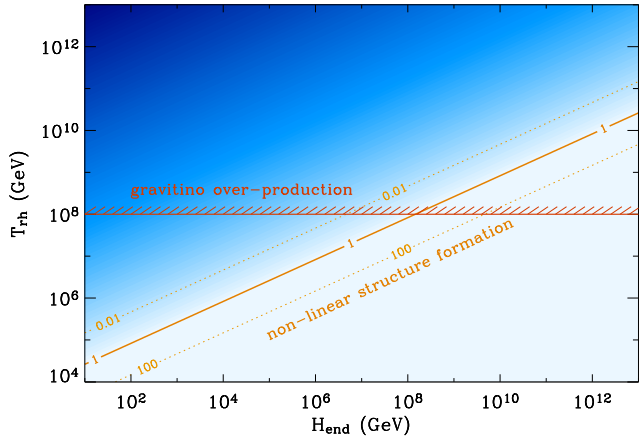}
  \caption[...]{Contour plot of the r.m.s. density fluctuation
    $\delta\rho _{\bm k}/\rho$ on the scale corresponding to the
    horizon at the end of inflation as measured at the onset of the
    radiation era in the $T_{\rm rh}$ (reheating temperature)--$H_{\rm
      end}$ (Hubble constant at the end of inflation). The solid line
    that is labelled ``$1$'' delineates parameter space where
    non-linear structures form between the end of inflation and
    reheating (the region below this line) from parameter space where
    the Universe remains in the linear regime of density
    perturbations, \ie $\delta\rho _{\bm k}/\rho \lsim 1$, before
    reheating (the blue region above the line). The normalization for
    the curvature perturbations on the Hubble scale at the end of
    inflation has been assumed to be ${\cal P}_{\zeta}\sim 10^{-11}$
    in accordance with Fig.~\ref{fig:primspectre}. The two dotted
    lines indicate where the power spectrum of the fluctuations is
    larger and smaller than one or, equivalently, how the solid line
    would move when the normalization of the curvature perturbations
    on the Hubble scale after inflation is higher, ${\cal
      P}_\zeta=10^{-9}$ (line labeled $0.01$), or lower, ${\cal
      P}_\zeta=10^{-13}$ (line labeled $100$). The solid red line
    shows approximate upper limits on $T_{\rm rh}$ in order to avoid
    gravitino overproduction. Note that the figure is based on
    Eq.~(\ref{eq:nl}), and makes the fairly generic assumptions that
    energy density after inflation is dominated by a single
    oscillating scalar field with potential $m^2\phi^2/2$, and that
    $\phi\lsim M_{_{\rm Pl}}$ at the end of inflation.}
\label{fig:TRH}
\end{figure}

Whether the Universe indeed entered an era with non-linear structures,
already before the epoch of reheating, depends on the reheat
temperature and the scale of inflation. Attempting to keep the reheat
temperature low enough to avoid overproduction of gravitinos, \ie
$T_{\rm rh}\simle 3\times 10^7-3\times 10^8$GeV
~\cite{Kawasaki:2008qe,Bailly:2009pe} implies that essentially all
large-field inflationary models are followed by a phase of non-linear
structure formation. Figure~\ref{fig:TRH} provides an illustration of
this fact assuming a quadratic potential during the oscillating
phase. It is seen that even for $H_{\rm end}$ as low as $10^7$GeV
non-linearities typically develop unless $T_{\rm rh}\gsim
10^8$GeV. Therefore, a nonlinear cosmic phase before reheating appears
to be a generic prediction. Note, however, that these considerations
assume that after inflation energy density is dominated by a single
oscillating scalar field.

These findings may not be surprising since it is known that the
effective equation of state during such a phase is that of a Universe
filled with pressure-less dust. One would thus expect the inflaton
condensate to be gravitationally unstable, which indeed, has been
shown in the Newtonian limit~\cite{Khlopov:1985jw,Nambu:1989kh} and in
an averaged relativistic
limit~\cite{Ratra:1990me,Hwang:1996xd}. However, our analysis and the
effect that we report go beyond this Newtonian analogue, because this
instability is found to be intrinsically some form of metric
preheating. And, indeed, we find that a similar effect also takes
place in the case of a quartic inflaton potential, for which the
effective equation of state during oscillations is that of
radiation. In this latter case, the resonance is (and remains) sharply
peaked around the physical scale associated to the Hubble scale at the
end of inflation, but the growth of the density contrast is
exponential.

\par

The outline of the paper is as follows. In
Section~\ref{sec:generalconsiderations} the metric preheating of
scalar perturbations is analyzed in detail. The primordial power
spectrum is calculated numerically over a wide range of scales. In
Section~\ref{sec:mcarrephicarre}, we examine in detail the growth of
density perturbations in a quadratic inflaton potential, while in
Section~\ref{sec:quartic}, we discuss the case of a quartic inflaton
potential. Conclusions are drawn in Section~\ref{sec:conclusions}.

\section{General Considerations}
\label{sec:generalconsiderations}

The evolution of scalar (density) perturbations is controlled by
the so-called Mukhanov variable. If matter is described by a scalar
field (as it is the case during inflation and preheating), then the
equation of motion of the Fourier transform of Mukhanov's variable is
given
by~\cite{Mukhanov:1990me,Martin:2003bt,Martin:2004um,Martin:2007bw}
\begin{equation}
\label{eq:eqmotv}
v_{{\bm k}}''+\left[k^2-\frac{\left(a\sqrt{\epsilon_1}\right)''}
{a\sqrt{\epsilon_1}}\right]v_{{\bm k}}=0 \, .
\end{equation}
In this article a prime denotes a derivative with respect to conformal
time while a dot means derivative with respect to cosmic time. The
quantity $k$ is the comoving wave number and $\epsilon _1\equiv
-\dot{H}/H^2$ is the first slow-roll parameter. The quantity
$H=\dot{a}/a$ is the Hubble parameter and $a$ denotes the
Friedman-Lema\^{\i}tre-Robertson-Walker scale factor. Mukhanov's
variable is related to the curvature perturbation, 
\begin{equation}
\label{eq:zetavsv}
\zeta_{\bm k}=\sqrt{\frac{\kappa}{2}}\frac{v_{\bm k}}{a\sqrt{\epsilon
    _1}}\ ,
\end{equation}
where $\kappa \equiv 8\pi/\mpl^2$, $\mpl$ being the
(non-reduced) Planck mass. The importance of the curvature
perturbation lies in the fact that, on large scales, it is a conserved
quantity~\cite{Martin:1997zd} and, therefore, it can be used to
propagate the inflationary spectrum from the end of inflation to the
post-inflationary era. As a consequence of the above relation, the
spectrum of $\zeta _{\bm k}$ can be expressed as
\begin{equation}
{\cal P}_{\zeta}(k)\equiv \frac{k^3}{2\pi ^2}
\left\vert \zeta_{\bm k}\right\vert ^2
=\frac{2k^3}{\pi \mpl^2}\left\vert 
\frac{v_{\bm k}}{a\sqrt{\epsilon _1}}\right\vert ^2\, .
\label{Pzeta}
\end{equation}
In order to calculate ${\cal P}_{\zeta}(k)$, one needs to integrate
Eq.~(\ref{eq:eqmotv}), which requires the knowledge of the initial
conditions for the mode function $v_{\bm k}$. Since, at the
beginning of inflation, all the modes of astrophysical interest today
were much smaller than the Hubble radius, the initial condition are
chosen to be the Bunch-Davis vacuum which amounts to
\begin{equation}
\lim_{k/(aH)\rightarrow +\infty}v_{\bm k}=\frac{1}{\sqrt{2k}}
{\rm e}^{-ik\eta }\, ,
\label{eq:initial}
\end{equation}
where $\eta $ denotes conformal time.

\par

The above formulation is well-adapted to the calculation of the power
spectrum during inflation. However, in order to study the metric
perturbations during the phase where the inflaton field oscillates it
is more convenient to work in terms of cosmic time and to use the
rescaled variable defined by $\tilde{v}_{\bm k}=a^{1/2}v_{\bm
  k}$. Then, Eq.~(\ref{eq:eqmotv}) takes the
form~\cite{Finelli:1998bu}
\begin{eqnarray}
\ddot{\tilde{v}}_{\bm k} &+& \Biggl[\frac{k^2}{a^2}
+\frac{{\rm d}^2V}{{\rm d}\phi ^2}
+3\kappa \dot{\phi}^2
-\frac{\kappa ^2}{2H^2}\dot{\phi}^4+\frac{3\kappa }{4}
\left(\frac{\dot{\phi}^2}{2}-V\right)
\nonumber \\ &+& 
2\kappa \frac{\dot{\phi}}{H}\frac{{\rm d}V}{{\rm d}\phi}\Biggr]
\tilde{v}_{\bm k}=0\, .
\label{eq:evolution}
\end{eqnarray}
This formulation is particularly convenient as all the terms in the
above equation remain non-singular even when the field oscillates at
the bottom of its potential, that is to say when $\phi $, $\dot{\phi}$
and $\epsilon_1$ periodically
vanish~\cite{Kodama:1996jh,Hamazaki:1996ir}.

\par

To proceed further, one needs to choose the potential $V(\phi)$. The
form of the inflaton potential is presently unknown (although it is
constrained, see
Refs.~\cite{Martin:2006rs,Lorenz:2007ze,Lorenz:2008je}). A convenient
way to treat the most general case is to consider the following
potential
\begin{equation}
\label{eq:pot}
V(\phi)=V_0 \pm \frac{1}{2}m^2\phi^2+\frac{\lambda }{4}\phi^4+\cdots \, ,
\end{equation}
where the dots refer to higher order terms. As it is well-known, this
expansion can be difficult to control, most notably so for large field
models where, during inflation, the vacuum expectation value of the
field is large compared to the Planck mass. However, one can always
assume that some terms in the above expansion are absent, for instance
due to particular symmetries of the underlying theory. Different
choices of coefficients in the above equation~(\ref{eq:pot})
correspond to different models. With the three terms displayed in
Eq.~(\ref{eq:pot}), one can describe large field models, small field
models and hybrid inflation (at least in the inflationary valley). In
the next section, we consider the massive case.

\bigskip

\section{The Quadratic Case}
\label{sec:mcarrephicarre} 

\subsection{Density Perturbations}
\label{subsec:quadraticdp}

In this section, we investigate a model with a simple quadratic
potential,
\begin{equation}
\label{eq:quadraticpot}
V(\phi)=\frac{1}{2}m^2\phi^2 \, .
\end{equation}
We have in mind a large field inflationary model, but as discussed
above, this quadratic potential has a more general applicability when
it comes to discussing the physics of inflaton oscillations after the
actual era of inflation. It may also describe the physics of curvaton
oscillations. We will therefore keep wherever needed, the
dependence of our results on the mass $m$ (and Hubble scale $H_{\rm
  end}$ at the end of inflation), although we will use as fiducial
values those corresponding to a large field inflationary model.

\par

In large field models, the mass $m$ is fixed by the normalization of
the amplitude of density fluctuations to that measured by COBE and
WMAP. Standard calculations lead to
\begin{equation}
\label{eq:mnormalized}
m\simeq \frac{\sqrt{180 \pi}}{2N_*+1}\frac{Q_{\rm rms-PS}}{T}\mpl\, 
\end{equation}
where $N_*$ is the number of e-folds before the end of inflation at
which the scale of astrophysical interest left the Hubble radius
during inflation (we take $N_*=50$) and where $Q_{\rm rms-PS}/T\simeq
6\times 10^{-6}$. This gives $m\simeq 1.4 \times 10^{-6} \mpl$.

\par

The power spectrum ${\cal P}_\zeta(k)$ of density fluctuations at the
end of inflation is calculated numerically in the following, via the
integration of Eq.~(\ref{eq:eqmotv}) throughout the inflationary
era. Although one may find excellent approximations to this power
spectrum on CMBR scales through the use of standard slow-roll
calculations, we indeed find that up to second order in the slow-roll
parameters, these methods cannot predict reliably the power spectrum
on the shorter scales that we are interested in. More specifically,
one finds to first order in the slow-roll
parameter~\cite{Leach:2002ar}

\begin{figure}
  \centering
  \includegraphics[width=0.5\textwidth,clip=true]{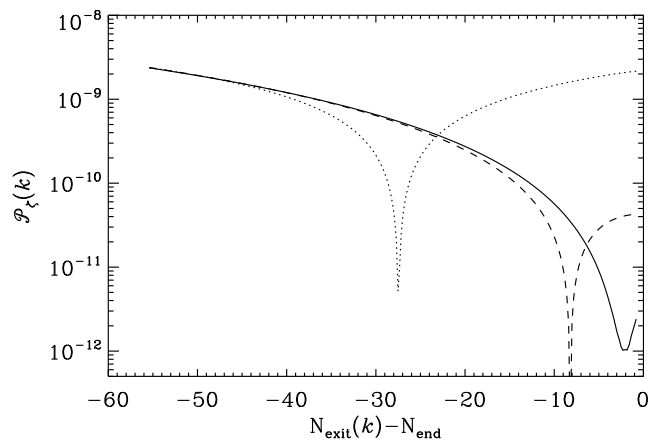}
  \caption[...]{Primordial power spectrum of curvature perturbations
    (solid line) for the large field model $V(\phi)=m^2\phi^2/2$. Here
    $N_{\rm exit}(k)-N_{\rm end}$ is ${\rm ln}\,[a_{\rm
      cross}(k)/a_{\rm end}]$, with $a_{\rm cross}(k)$ the scale
    factor at Hubble radius crossing of mode $k$ during inflation
    (i.e. $k/a_{\rm cross} \equiv 2\pi H$), and $a_{\rm end}$ the
    scale factor at the end of inflation. The dotted and dashed lines
    respectively represents the first and second order slow-roll
    results as given by Eqs.~(\ref{eq:sr1}) and~(\ref{eq:sr2}) with
    $\epsilon_1\simeq 1/[2(N_*+1/2)]$,
    $\epsilon_2=\epsilon_3=1/(N_*+1/2)$ and $N_*\simeq 50$. The
    slow-roll result is accurate in the CMBR region but breaks down on
    smaller scales where the predicted power spectra become
    negative. The upturn is related to modes that have not crossed out
    the Hubble radius and have remained in their vacuum state, see
    text for details.}
\label{fig:primspectre}
\end{figure}

\begin{widetext}
\begin{equation}
\label{eq:sr1}
{\cal P}_{\zeta}(k)=\frac{H^2}{\pi \epsilon_1\mpl^2}
\biggl[1-2\left(C+1\right)\epsilon_1-C\epsilon_2
+\left(-2\epsilon_1-\epsilon_2\right)\ln \left(\frac{k}{k_*}\right)\biggr]\, ,
\end{equation}
while, at second order, it is given by~\cite{Leach:2002ar}
\begin{eqnarray}
\label{eq:sr2}
{\cal P}_{\zeta}(k)&=&\frac{H^2}{\pi \epsilon_1\mpl^2}
\biggl\{1-2\left(C+1\right)\epsilon_1-C\epsilon_2
+\left(2C^2+2C+\frac{\pi^2}{2}-5\right)\epsilon_1^2
+\left(C^2-C+\frac{7\pi^2}{12}-7\right)\epsilon_1\epsilon_2
\nonumber \\ & &
+\left(\frac{C^2}{2}+\frac{\pi^2}{8}-1\right)\epsilon_2^2
+\left(-\frac{C^2}{2}+\frac{\pi ^2}{24}\right)\epsilon_2\epsilon_3
+\left[-2\epsilon_1-\epsilon_2+2\left(2C+1\right)\epsilon_1^2
+\left(2C-1\right)\epsilon_1\epsilon_2+C\epsilon_2^2-C\epsilon_2\epsilon_3
\right]
\nonumber \\ & & \times
\ln\left(\frac{k}{k_*}\right)
+\frac{1}{2}\left(4\epsilon_1^2+2\epsilon_1\epsilon_2
+\epsilon_2^2-\epsilon_2\epsilon_3
\right)\ln ^2\left(\frac{k}{k_*}\right)\biggr\}\, .
\end{eqnarray}
\end{widetext}
In these expressions, $\epsilon_2$ and $\epsilon_3$ are slow-roll
parameters belonging to the Hubble flow
hierarchy~\cite{Schwarz:2001vv,Leach:2002ar} $\epsilon_{n+1}\equiv
{\rm d}\ln \epsilon_n/({\rm d}N)$ ($N$ being the number of e-folds)
and $C$ is a numerical constant $C\simeq -0.7296$. The quantity $k_*$
is the so-called pivot scale the location of which is arbitrary but is
usually chosen in the middle of the range of scales probed by the
CMBR. These approximations fail to reproduce the power spectrum on
small scales because the lever arm is then too important for the
accuracy of the slow-roll calculation. This is rather unusual, as the
slow-roll method generally provides an excellent approximation. This
break-down of the slow-roll calculation on small scales is clearly
seen in Fig.~\ref{fig:primspectre}, which compares the full numerical
calculation of the power spectrum with the above approximations for a
large field model quadratic inflationary model.  As one might expect,
second-order slow-roll calculations are more accurate on small scales
than the first order (which, among others justifies the techniques
that have been developed for this
purpose~\cite{Gong:2001he,Martin:2002vn}).

\begin{figure}
  \centering
  \includegraphics[width=0.5\textwidth,clip=true]{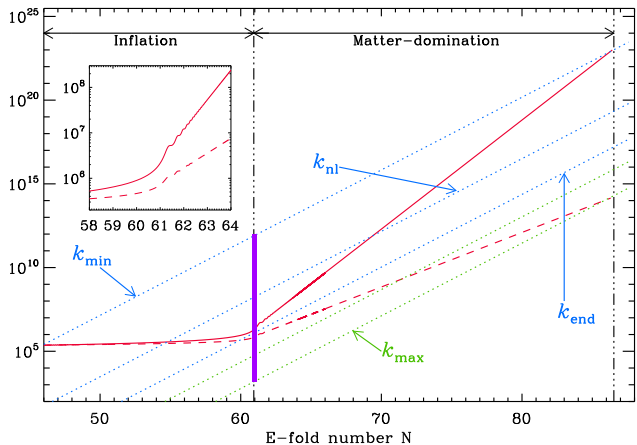}
  \caption[...]{Hubble radius $\ell _{_{\rm H}}$ (red solid line),
    lower bound of resonance band $\ell _{_{\rm C}}$ (red dashed line)
    and physical wavelengths (blue and green dotted lines), in Planck
    units, versus the e-fold number during the final stage of
    inflation and the inflaton oscillation era for a large field
    $m^2\phi^2/2$ model. The reheating temperature is $T_{\rm
      rh}=10^7\mbox{GeV}$ hence the number of e-folds during the
    inflaton oscillation era is $N_{\rm rh}\simeq 25.6$. The inset shows
    the detailed behaviors of $\ell _{_{\rm H}}$ and $\ell _{_{\rm
        C}}$ at the transition between inflation and inflaton oscillation.}
\label{fig:scaleinf}
\end{figure}

\subsubsection{Evolution of Density Perturbations after Inflation}

We are now interested in the behavior of perturbations after the end
of inflation.  On time scales $t\,\gg\,H_{\rm end}^{-1}$ (the
subscript ``${\rm end}$'' denotes hereafter quantities evaluated at
the end of inflation), the inflaton vacuum expectation value evolves
approximately as
\begin{equation}
\label{eq:inflatonosci}
\phi (t)\simeq \phi_{\rm end}\left(\frac{a_{\rm end}}{a}
\right)^{3/2}\sin\left(mt\right)\, ,
\end{equation}
for a quadratic inflaton potential. Let us recall that $\phi_{\rm
  end}\simeq \mpl/(2\sqrt{\pi})$ and $H_{\rm end}\simeq m/\sqrt{3}$ in the
particular case of a large field quadratic inflationary model.
Plugging Eq.~(\ref{eq:inflatonosci}) into Eq.~(\ref{eq:evolution}), it
is easy to see that the dominant term of the last four gravitational
terms is the last one as it scales as $1/a^{3/2}$ while the others
decay as $1/a^3$. Neglecting the other three terms one finds an
equation of motion of the form
\begin{equation}
\frac{{\rm d}^2\tilde{v}_{\bm k}}{{\rm d}z^2}
+\left[1+\frac{k^2}{m^2a^2}
-\sqrt{6\kappa}\phi_{\rm end}\left(\frac{a_{\rm end}}{a}
\right)^{3/2}\cos\left(2z\right)\right]\tilde{v}_{\bm k}=0\, ,
\end{equation}
where we have defined $z\equiv mt+\pi/4$. This equation is similar to 
a Mathieu equation 
\begin{equation}
\frac{{\rm d}^2\tilde{v}_{\bm k}}{{\rm d}z^2} +
\bigl[A_{\bm k}-2q{\rm cos}(2z)\bigr]\tilde{v}_{\bm k} = 0
\end{equation}
with 
\begin{eqnarray}
\label{eq:a}
A_{\bm k} &=& 1+\frac{k^2}{m^2a^2}\, ,\\
\label{eq:q}
q &=& \frac{\sqrt{6\kappa}}{2}\phi_{\rm end}\left(\frac{a_{\rm end}}{a}
\right)^{3/2}\, .
\end{eqnarray}
Since $q\ll 1$, we are in the narrow resonance regime. The first
instability band is given by $1-q<A_{{\bm k}}<1+q$ which amounts to
\begin{equation}
\label{eq:defband}
0<\frac{k}{a}<\sqrt{3Hm}\, .
\end{equation}

From the above inequality, we note the emergence of a new
characteristic spatial length given by $\ell _{_{\rm C}}\equiv
1/\sqrt{3Hm}$, which sets the lower bound to the range of resonant
spatial scales. During inflation, in a large field model, $\ell
_{_{\rm C}}$ is almost a constant and $\ell_{_{\rm C}}\sim\ell_{_{\rm
    H}}$, with $\ell_{_{\rm H}}\equiv H^{-1}$.  After the end of
inflation, during the inflaton oscillation era, with $H\propto
a^{-3/2}$ one has $\ell _{_{\rm C}}\propto a^{3/4}$ and, therefore,
$\ell _{_{\rm C}}$ grows more slowly than the Hubble radius $\ell
_{_{\rm H}}\propto a^{3/2}$. The situation is summarized in
Fig.~\ref{fig:scaleinf} where both scales are displayed (solid and
dashed red lines). Their evolution has been computed by an exact
numerical integration of the equations of motion in the case of a
large field model. Since the physical wavelength of a comoving Fourier
mode, $\lambda \propto a$, grows faster than $\ell _{_{\rm C}}$, the
scales may cross $\ell _{_{\rm C}}$ during inflation or during the
matter-dominated era and thereby enter the instability
band. Furthermore, a mode that has entered this band will always
remain in it, at least until reheating is completed. One can thus
distinguish two types of modes: those that have entered the resonance
band during inflation (blue lines in Fig.~\ref{fig:scaleinf}) and
those that have entered it during the inflaton oscillation era (green
lines in Fig.~\ref{fig:scaleinf}). Of course, the preheating
instability is only operative after inflation and before reheating.

\par

Having identified the scales affected by the resonance, we now discuss
their amplification. For the first instability band, the Floquet index
is given by $\mu =q/2$. Then the mode function evolves as
$\tilde{v}_{\bm k}\propto {\rm e}^{\mu z}$. However, in the present
case, we have a time-dependent Floquet index, for which the
corresponding solution can be written as~\cite{Finelli:1998bu}
\begin{equation}
\tilde{v}_{\bm k}
\propto \exp\left(\int \mu {\rm d}z\right)\propto a^{3/2}\, .
\end{equation}
Note that, taking the integral of the time-dependent $\mu$ does not
correspond to the WKB approximation, but rather is an educated guess
confirmed by numerical simulation, see below. This implies that
$v_{{\bm k}}\propto a$ and therefore $\zeta_{\bm k}$ is constant
according to its definition Eq.~(\ref{eq:zetavsv}). Note also that the
curvature perturbation usually remains constant only on super-Hubble
scales while decaying on sub-Hubble scales, unless there is
gravitational instability.

\par

\begin{figure}
  \centering
  \includegraphics[width=0.5\textwidth,clip=true]{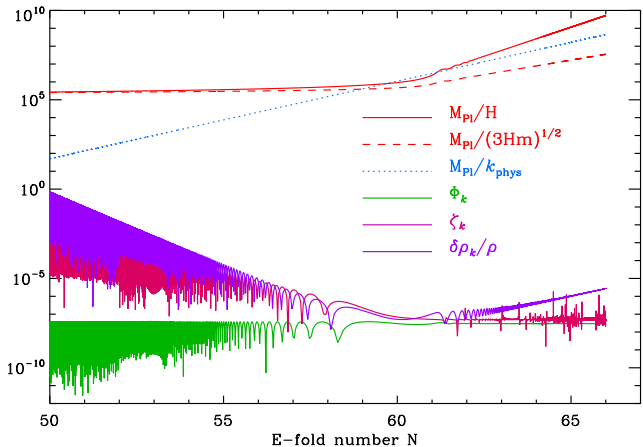}
  \caption[...]{Evolution of the density perturbations
    $\delta\rho_{\bm k}/\rho$, $\zeta _{\bm k}$ and $\Phi_{\bm k}$
    towards the end of inflation and during the ensuing period of
    metric preheating after inflation. The notation is as in the text
    with $k_{\rm phys}\equiv k/a$; $N$ denotes the e-fold number of
    the scale factor. The reduced Planck mass $M_{\rm Pl}\equiv
    \mpl/\sqrt{8\pi}$ is used in this plot.}
\label{fig:modpert}
\end{figure}

We are now interested in the behavior of the fractional mass-energy
density perturbation, $\delta_{{\bm k}}(\eta)\,\equiv\,\delta
\rho_{{\bm k}}/\rho$, where $\rho $ is the background energy density
of the scalar field. The perturbed Einstein equation implies that
\begin{equation}
\label{eq:deltak}
\delta_{{\bm k}}=-\frac{2}{3}
\left(\frac{k^2}{a^2H^2}+3\right)\Phi_{{\bm k}}
-2\frac{\dot{\Phi}_{\bm k}}{H}\, ,
\end{equation}
where $\Phi_{\bm k}$ is the Bardeen
potential~\cite{Bardeen:1980kt,Mukhanov:1990me}. The Bardeen potential
is related to the quantity $\zeta _{\bm k}$ by
\begin{equation}
\label{eq:defzeta}
\zeta_{\bm k}=\frac{2}{3}\frac{H^{-1}\dot{\Phi}_{\bm k}
+\Phi_{\bm k}}{1+p/\rho}+\Phi_{\bm k}\, ,
\end{equation}
where $p$ denotes the pressure. For constant $p/\rho$, if $\Phi _{\bm
  k}$ is constant so is $\zeta _{\bm k}$. This is a generic result on
large scales in one fluid models. If $\Phi_{\bm k}$ evolves as a
power-law in the scale factor (also a generic result), $\Phi_{\bm
  k}\,\propto\,a^s$, then $H^{-1}\dot{\Phi}_{\bm k} = s \Phi_{\bm k}$
and consequently, Eq.~(\ref{eq:defzeta}) indicates that $\zeta_{\bm
  k}$ follows the same power-law behavior. Therefore, the constancy of
$\zeta_{\bm k}$ in the resonance band implies that $\Phi_{\bm k}$ also
remains constant.  This is confirmed by a numerical integration of the
equations of motion, see Fig.~\ref{fig:modpert}. Then $\zeta_{\bm
  k}=5\Phi_{\bm k}/3$, as the pressure vanishes on average during the
preheating stage. Consequently, one may express $\delta_{\bm k}$ as a
function of $\zeta_{\bm k}$ as follows
\begin{equation}
\label{eq:deltakvszeta}
\delta_{{\bm k}}=-\frac{2}{5}
\left(\frac{k^2}{a^2H^2}+3\right)\zeta_{{\bm k}}\, .
\end{equation}
This relation remains valid in both the large scale and the small
scale limit (with respect to the Hubble radius) provided the modes are
in the instability band (i.e. $\zeta_{\bm k}$ is constant). Given that
$a H\propto a^{-1/2}$ in a inflaton oscillation dominated Universe,
one then finds that $\delta_{{\bm k}}$ grows as $a(t)$ on sub-Hubble
scales but remains constant on super-Hubble scales. This behavior is
also confirmed by the numerical integration whose results are shown in
Fig.~\ref{fig:modpert}. The preheating instability for all sub-Hubble
modes with $k/a <\sqrt{3Hm}$ may thus be understood as the
gravitational instability of the oscillating scalar condensate.

\par

Let us summarize our main findings. We have found that all modes in
the instability band and outside the Hubble radius are such that
$\zeta _{{\bm k}}$ and $\delta_{{\bm k}}$ are constant. This is of
course the case for modes on scales probed by the CMBR and, therefore,
one recovers the fact that the inflationary power spectrum propagates
through the reheating stage without being affected. However, for those
modes which are in the instability band and inside the Hubble radius,
one finds that $\zeta _{{\bm k}}$ is constant but that $\delta_{{\bm
    k}}$ grows as $a(t)$. Therefore, and contrary to the standard
lore, the preheating stage can affect the behavior of the
perturbations even if there is only one non-interacting fluid.

\subsubsection{Spectrum of Density Perturbations at Reheating}

From the above considerations, we see that the amplification described
above occurs only for a range of scales, $k\in \left[k_{\rm
    min},\,k_{\rm max}\right]$ in terms of comoving wavenumber. We now
discuss this point in more detail and specify $k_{\rm min}$ and
$k_{\rm max}$. Henceforth, we introduce the short-hand notation:
\begin{equation}
  \hat k\,\equiv\, \frac{k}{a_{\rm end}H_{\rm end}}\ ,\label{eq:khat}
\end{equation}
where the label ``end'' refers as before to the end of inflation.  The
smallest unstable spatial scale is that which enters the resonance
band at the time of completion of reheating.  It corresponds to the
green dotted line labelled ``$k_{\rm max}$'' in
Fig.~\ref{fig:scaleinf}. Interestingly enough, this scale has never
been super-Hubble. From the condition $k_{\rm max}/a_{\rm
  rh}=\sqrt{3H_{\rm rh}m}$, where $H_{\rm rh}$ is the Hubble parameter
at the end of the reheating epoch or at the beginning of the radiation
dominated phase, one obtains
\begin{eqnarray}
\label{eq:kmax}
  \hat{k}_{\rm max} &=& 1.37\times 10^3
\left(\frac{m}{1.4\times 10^{-6}\mpl}\right)^{1/2} 
\nonumber \\ & & \times
\left(\frac{T_{\rm rh}}{10^7\mbox{GeV}}\right)^{-1/3}
\, \left(\frac{H_{\rm end}}{10^{13}\mbox{GeV}}\right)^{-1/3}\, .
\end{eqnarray}
The last equation assumes $g_*=230$ degrees of freedom at
reheating. All modes with $\hat k< \hat k_{\rm max}$ have entered the
resonance band during inflation or during the preheating era.

\par

We have also seen that a necessary condition for the growth of
$\delta_{\bm k}$ on a given scale is that the corresponding mode lies
within the Hubble radius at some time before the reheating epoch. As a
consequence, the largest spatial scale that is unstable with respect
to the growth of $\delta_{\bm k}$ is that which re-enters the Hubble
radius at the end of the pre-reheating epoch, i.e. $k_{\rm min}\simeq
a_{\rm rh}H_{\rm rh}$. This corresponds to the blue dotted line
labeled ``$k_{\rm min}$'' in Fig.~\ref{fig:scaleinf} and
\begin{eqnarray}
\label{eq:kmin}
  \hat{k}_{\rm min} &=& 2.74\times 10^{-6}
\left(\frac{T_{\rm rh}}{10^7\mbox{GeV}}\right)^{2/3}
\, \left(\frac{H_{\rm end}}{10^{13}\mbox{GeV}}\right)^{-1/3}\, .
\end{eqnarray}
As the density contrast increases in direct proportion to the scale
factor for modes in the range $k\in \left[k_{\rm min},\,k_{\rm
    max}\right]$, non-linear structures may eventually form. The
amplitude of the density contrast at the end of the preheating epoch
is given by the density contrast at the end of inflation and by the
e-fold number of growth while the mode is sub-horizon sized and in the
resonance band.  At this point, one should distinguish the modes that
enter the resonance band during inflation (the dotted blue lines in
Fig.~\ref{fig:scaleinf}) and those that have entered the band during
the preheating era but that have never exited the Hubble radius (the
dotted green lines in Fig.~\ref{fig:scaleinf}).

\par

We first address the former ones, for which $k\in \left[k_{\rm
    min},\,k_{\rm end}\right]$, with $\hat k_{\rm end}\equiv 1$. Given
that, in the resonance band, the Fourier density contrast remains
constant on super-Hubble scales, the behavior of $\delta _{\bm k}$ on
sub-Hubble scale can be approximated as
\begin{equation}
\delta _{\bm k}=\delta _{\bm k}^{\rm hc}\frac{a}{a_{\rm hc}}
=\delta _{\bm k}^{\rm end}\frac{a}{a_{\rm hc}}\, ,
\end{equation}
where the label ``${\rm hc}$'' refers to the epoch of Hubble radius
crossing of mode ${\bm k}$ and $\delta _{\bm k}^{\rm end}$ is the
density contrast calculated at the end of inflation, see
Fig.~\ref{fig:primspectre}.  Let us precise the value of $a_{\rm hc}$:
in order to match both asymptotic behaviors of $\delta_{\bm k}$ in the
super-horizon, namely $\delta_{{\bm k}} \rightarrow 6\zeta_{{\bm
    k}}/5$ and in the sub-horizon regimes, namely $\delta_{{\bm
    k}}\rightarrow \left[k/(aH)\right]^22\zeta_{{\bm k}}/5$, one
defines $a_{\rm hc}$ as that at which $ k = \sqrt{3} a_{\rm hc}H_{\rm
  hc}$. This implies that $a_{\rm hc} = 3 a_{\rm end} \hat
k^{-2}$. Now, using Eq.~(\ref{eq:deltakvszeta}), one can write $\delta
_{\bm k}^{\rm end}=(6/5)\sqrt{2\pi^2 {\cal P}_{\zeta}(k)/k^3}$. Then,
the power spectrum of $\delta _{\bm k}$ at the time of reheating is
evaluated as
\begin{eqnarray}
\label{eq:powerspectrumband}
{\cal P}_{\delta }^{\rm rh}(k)&=&\frac{k^3}{2\pi^2} 
\left \vert \delta _{\bm k}^{\rm end}\right\vert^2
\left(\frac{a_{\rm rh}}{a_{\rm hc}}\right)^2
\nonumber \\ 
&=& 2.82 \times 10^{10}\, \hat{k}^4
\left(\frac{{\cal P}_{\zeta}}{10^{-11}}\right)\,
\left(\frac{T_{\rm rh}}{10^7\mbox{GeV}}\right)^{-8/3}
\nonumber \\ & & 
\times \left(\frac{H_{\rm end}}{10^{13}\mbox{GeV}}\right)^{4/3}\, .
\label{eq:nl}
\end{eqnarray}
The variance of the density fluctuation enclosing the scale defined by
$k$ is directly expressed at reheating as 
\begin{equation}
\langle \delta ^2\rangle_k =\int_0^k \,\,{\rm d}\,{\rm ln}\,k'\,\,{\cal
P}_\delta^{\rm rh}(k') \ ,
\end{equation}
and the condition $\langle \delta ^2\rangle_k \simeq 1$ thus roughly
amounts to ${\cal P}_{\delta}^{\rm rh} \simeq 1$. This singles out a
specific spatial scale (corresponding to wavenumber $k_{\rm nl}$),
such that scales with $k< k_{\rm nl}$ have not had sufficient time to
become non-linear, while those with $k>k_{\rm nl}$ have become
non-linear by reheating. Straightforward manipulations lead to
\begin{eqnarray}
\label{eq:defknl}
\hat{k}_{\rm nl}& \simeq & 4.50\times 10^{-3}
\left(\frac{{\cal P}_{\zeta}}{10^{-11}}\right)^{-1/4}\,
\left(\frac{T_{\rm rh}}{10^7\mbox{GeV}}\right)^{2/3}
\nonumber \\ & & 
\times \left(\frac{H_{\rm end}}{10^{13}\mbox{GeV}}\right)^{-1/3}\, .
\end{eqnarray}

Let us now address the unstable modes that have always been
sub-Hubble, namely $k\in \left[k_{\rm end},\,k_{\rm max}\right]$. For
these modes, Mukhanov's variable is still given by the vacuum
solution, Eq.~(\ref{eq:initial}), until it enters the instability
band. This leads to
\begin{equation}
\zeta_{\bm k}=\frac{{\rm e}^{-ik(\eta -\eta _{\rm end})}}{\sqrt{\epsilon_1}}
\left(\frac{a_{\rm end}}{a}\right)\zeta _{\bm k}^{\rm end}\, ,
\end{equation}
where $\epsilon_1\sim 3/2$ in the preheating era. In order to derive
this expression, we have set $\epsilon_1 = 1$ at the end of inflation
and we have assumed that Mukhanov's variable has remained constant
through the (instantaneous) transition from inflation to the
preheating era; this is expected insofar as the modes are well
sub-Hubble at that time, $k\,\gg\,aH$, hence Mukhanov's variable
oscillates freely. The quantity $\zeta _{\bm k}^{\rm end}$ represents
the value of $\zeta _{\bm k}$ at the end of inflation or at the onset
of the oscillatory phase.  Using Eq.~(\ref{eq:defzeta}) with $p=0$,
one finds $\zeta _{\bm k}=\Phi_{\bm k}$ since $\zeta_{\bm k}$ and
$\Phi_{\bm k}$ must share the same power-law behavior $\propto a^{-1}$
(as discussed above), hence $H^{-1}\dot{\Phi}_{\bm k}=-\Phi_{\bm
  k}$. Then, Eq.~(\ref{eq:deltak}) leads to
\begin{eqnarray}
\label{eq:densitysmallscales}
\delta _{\bm k}&=&-\frac{2}{3}\frac{k^2}{a^2H^2}\Phi_{\bm k}\,\nonumber\\
&=&-\frac{2\hat{k}^2}{3}\frac{{\rm e}^{-ik(\eta -\eta _{\rm end})}}
{\sqrt{\epsilon_1}}\zeta _{\bm k}^{\rm end}\,\nonumber\\
&=&C(k){\rm e}^{-ik\eta}\, .
\end{eqnarray}
The density contrast then oscillates with a constant amplitude until
the mode enters the resonance band, in which $\delta
_{\bm k}=C(k)a/a_{\rm rc}$ where $a_{\rm rc}$ denotes the scale
factor at the entry into the resonance band. This latter is given by
$a_{\rm rc}=k^4/(9m^2a_{\rm end}^3H_{\rm end}^2)$ which implies that
$a_{\rm rh}/a_{\rm rc}=(\hat{k}_{\rm max}/\hat{k})^4$. The power
spectrum of density fluctuations at the time of preheating can now be
computed as
\begin{eqnarray}
\label{eq:spectresmallscales}
{\cal P}_{\delta }^{\rm rh}(k)&=& \frac{8\hat{k}^4}{27}
\left(\frac{\hat{k}_{\rm max}}{\hat{k}}\right)^8{\cal P}^{\rm end}_{\zeta}(k)
\, .
\end{eqnarray}
The quantity ${\cal P}^{\rm end}_{\zeta}$ is the power spectrum at the end of
inflation for the modes that have never exited the Hubble radius,
\begin{equation}
{\cal P}^{\rm end}_{\zeta}(k)=
\frac{\hat{k}^2}{\pi}\left(\frac{H_{\rm end}}{\mpl}\right)^2\, .
\end{equation}
We can check that it gives the correct order of magnitude by comparing
with the exact result in Fig.~\ref{fig:primspectre} for the case of a
large field model. Inserting this expression in
Eq.~(\ref{eq:spectresmallscales}), we see that the spectrum now
decreases as $\hat{k}^{-2}$ while it was proportional to $\hat{k}^4$
in the other range of wavenumbers $[k_{\rm min}, k_{\rm end}]$ that we
discussed previously.

\par

Finally, for those modes that have never entered the resonance band
and never crossed out the Hubble radius, that is to say
$\hat{k}>\hat{k}_{\rm max}$, the density contrast always behaves
according to Eq.~(\ref{eq:densitysmallscales}) and, therefore, the
corresponding power spectrum is given by
\begin{eqnarray}
\label{eq:spectreultrasmallscales}
{\cal P}_{\delta }^{\rm rh}(k)&=& \frac{8\hat{k}^4}{27}{\cal P}_{\zeta}(k)
\, .
\end{eqnarray}
This spectrum is proportional to $k^6$. Of course, these modes are
still in a vacuum state and therefore, despite the fact that the
vacuum fluctuations are large (which is likely to be linked to the
pathological behavior of the theory in the ultra-violet regime), there
is no energy density associated to this phenomenon. One way to see
this is to compute the evolution of the modes, and in particular the
Bogoliubov coefficient $\beta _{\bm k}$ of the negative frequency WKB
branch at late times, on sub-horizon scales (assuming the vacuum
corresponds to the positive frequency branch). For these vacuum modes,
$\beta _{\bm k}=0$ because the WKB approximation has always remained
valid. The energy density stored in these modes, which is given by
\begin{equation}
  \frac{{\rm d}\delta \rho _{\bm k}}{{\rm d}\ln \hat{k}}
\simeq \frac{k^4}{2\pi ^2a^4}\vert \beta _{\bm k}\vert ^2\, ,
\end{equation}
consequently vanish.

\begin{figure}
  \centering
  \includegraphics[width=0.5\textwidth,clip=true]{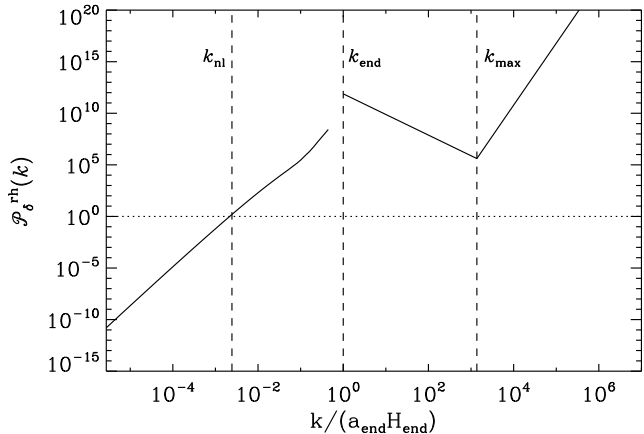}
  \caption[...]{Power Spectrum ${\cal P}_{\delta}^{\rm rh}(k)$,
    computed at reheating as given by
    Eqs.~(\ref{eq:powerspectrumband}),~(\ref{eq:densitysmallscales})
    and~(\ref{eq:spectreultrasmallscales}). In the range $k\in [k_{\rm
      min},k_{\rm end}]$, the spectrum is evaluated numerically, see
    Fig.~\ref{fig:primspectre}. The small discontinuity observed at
    $k_{\rm end}$ is just a numerical artefact, the actual power
    spectrum being of course continuous.}
\label{fig:specreheat}
\end{figure}

The situation is summarized in Fig.~\ref{fig:primspectre} where the
spectrum ${\cal P}_{\delta}(k)$ is displayed in the different regimes
studied before. The striking feature of this plot is of course that,
for $k>k_{\rm nl}$, the power spectrum is non-linear at the time of
reheating; let us stress that of course, the plotted values only
represent the extrapolation of the linear calculation.

\par

The non-linearity of modes with $k>k_{\rm end}$ is rather intriguing,
as these modes have never crossed out the Hubble radius, hence one
might question their classical nature. However, it is rather
straightforward to show that for these modes, $\vert\beta_{\bm
  k}\vert\sim 1$, hence energy density is indeed stored on those
scales. From this point of view, the non-linearities observed in the
power spectrum should a priori be taken seriously and viewed as a sign
that structure formation has started on those scales. In order to
assess the nature of these modes, one should apply the discussion of
the so-called quantum to classical transition of
modes~\cite{Lesgourgues:1996jc,Martin:2007bw} to the present
case. However, this question is beyond the scope of this paper.

\subsubsection{Associated Mass Scales}

Finally, it is interesting to quantify the range of masses associated
with the collapsing structures. In terms of the wavenumber of the
perturbation, the enclosed mass can be written as
\begin{equation}
M=\frac{4\pi }{3}(2\pi)^3\frac{\rho_{\rm end}}{H_{\rm end}^3}\hat{k}^{-3}\, .
\end{equation}
According to the previous considerations, the mass corresponding to
maximum growth of the density fluctuations corresponds to the scale
$k_{\rm end}$. Using the above expression, straightforward
manipulations lead to the following expression
\begin{equation}
M_{\rm end}\simeq 3.3\times10^3\left(\frac{H_{\rm end}}{10^{13}\mbox{GeV}}
\right)^{-1}\mbox{g}\, .
\end{equation}
The minimal mass, associated to $k_{\rm max}$ is given by 
\begin{eqnarray}
  M\left(k_{\rm max}\right)&\simeq& 1.3\times 10^{-6}\,
  \left(\frac{T_{\rm rh}}{10^7\mbox{GeV}}\right)\nonumber \\ & & \times 
  \left(\frac{m}{1.4\times 10^{-6}\mpl}\right)^{-3/2}\mbox{g}\, .
\end{eqnarray}
The maximum mass is associated to the scale $k_{\rm nl}$, i.e.
\begin{equation}
M_{\rm nl}\simeq 3.6\times 10^{10}
\left(\frac{{\cal P}_{\zeta}}{10^{-11}}\right)^{3/4}\,
\left(\frac{T_{\rm rh}}{10^7\mbox{GeV}}\right)^{-2}\mbox{g}\, .
\end{equation}
Quite interestingly, the abundance of primordial black holes with mass
$\lesssim 10^9\,$g is constrained by early Universe cosmology. Such
black holes indeed evaporate before BBN has started and the black hole
temperature is high enough to allow the production of copious amounts
of lightest supersymmetric particles, in excess of the closure
density~\cite{Green:1999yh}, as well as gravitinos and other late time
decaying massive particles, whose decay products would ruin the
success of BBN~\cite{Lemoine:2000sq,Khlopov:2004tn}. If such black
holes leave Planck-mass relics at their final point of evaporation,
additional constraints can be obtained from closure density
arguments~\cite{Carr:1994ar}. For black hole masses $\gtrsim 10^9\,$g,
stringent constraints can be derived from the impact of the evaporated
radiation on BBN, see ~\cite{Carr:2009jm} for a detailed discussion of
this and other observational constraints.

\subsection{Tensor modes}
\label{subsec:quadraticgw}

In the previous section, we have studied the amplification of density
perturbations during the reheating epoch.  Naturally, one may wonder
whether tensor modes can be amplified through a similar instability.
Their Fourier amplitude $h_{\bm k}=\mu_{\bm k}/a$ obeys
the equation~\cite{Grishchuk:1974ny}
\begin{equation}
\label{eq:motiongw}
\mu _{\bm k}''+\left(k^2-\frac{a''}{a}\right)\mu_{\bm k}=0\, .
\end{equation}
The effective potential is thus much simpler than in the case of
density perturbations as it only involves the scale factor and its
time derivatives up to second order. Defining $\bar{\mu}_{\bm k}\equiv
a^{1/2}\mu _{\bm k}$ and working in terms of the cosmic time, one
obtains
\begin{equation}
\label{eq:motiongw2}
\ddot{\bar{\mu}}_{\bm k}+\left(\frac{k^2}{a^2}-\frac{3\dot{H}}{2}
-\frac{9}{4}H^2\right)\bar{\mu}_{\bm k}=0\, .
\end{equation}
Using the explicit form of the inflaton field given by
Eq.~(\ref{eq:inflatonosci}), the equation of motion takes the form
\begin{equation}
  \frac{{\rm d}^2\bar{\mu}_{\bm k}}{{\rm d}z^2}
  +\left[\frac{k^2}{a^2m^2}+\frac{3\kappa}{8}\phi_{\rm end}^2
    \left(\frac{a_{\rm end}}{a}\right)^3\cos\left(2z\right)\right]
  \bar{\mu}_{\bm k}=0\, ,
\end{equation}
where $z\equiv mt$. Therefore, one also obtains a Mathieu equation
with
\begin{eqnarray}
\label{eq:agw}
A_{\bm k} &=& \frac{k^2}{a^2m^2}\, ,
\\
\label{eq:qgw}
q &=& -\frac{3\kappa}{16}\phi_{\rm end}^2
\left(\frac{a_{\rm end}}{a}\right)^3\, .
\end{eqnarray}
These equations should be compared to Eqs.~(\ref{eq:a})
and~(\ref{eq:q}). We are still in the regime $q\ll 1$ but the crucial
point is that there is no term $1$ in the expression of $A_{\bm k}$ in
Eq.~(\ref{eq:agw}). This means that, for a given scale, the condition
$1-q<A_{\bm k}<1+q$ cannot be maintained since $A_{\bm k}$ does not
scale as $q$ does. Therefore the amplification does not occur for
gravitational waves.

\section{The Quartic case}
\label{sec:quartic}

\subsection{A quartic potential $V(\phi) = \lambda \phi^4/4$}

\begin{figure*}
  \centering
  \begin{tabular}{cc}
  \includegraphics[width=0.45\textwidth,clip=true]{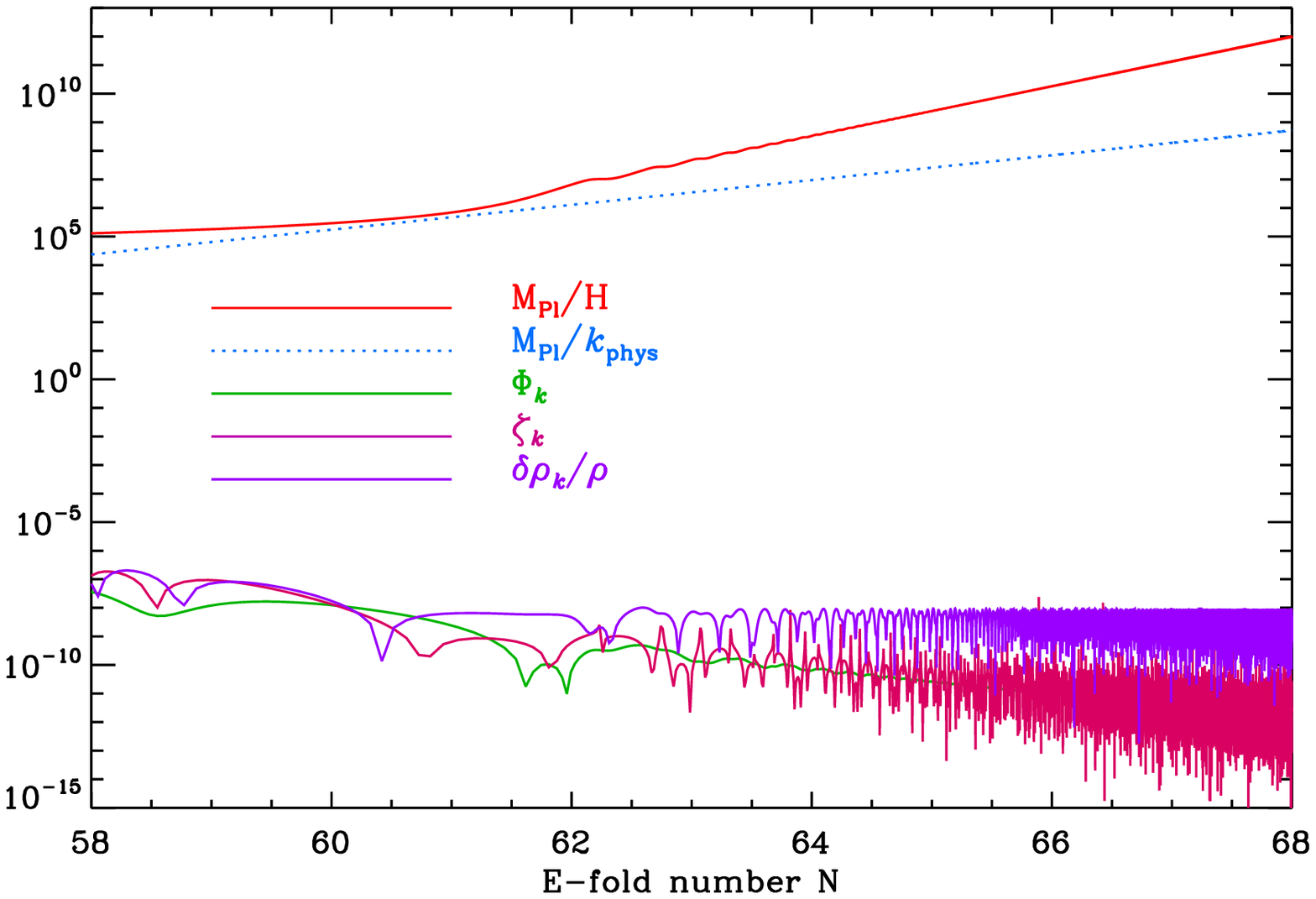} &
  \includegraphics[width=0.45\textwidth,clip=true]{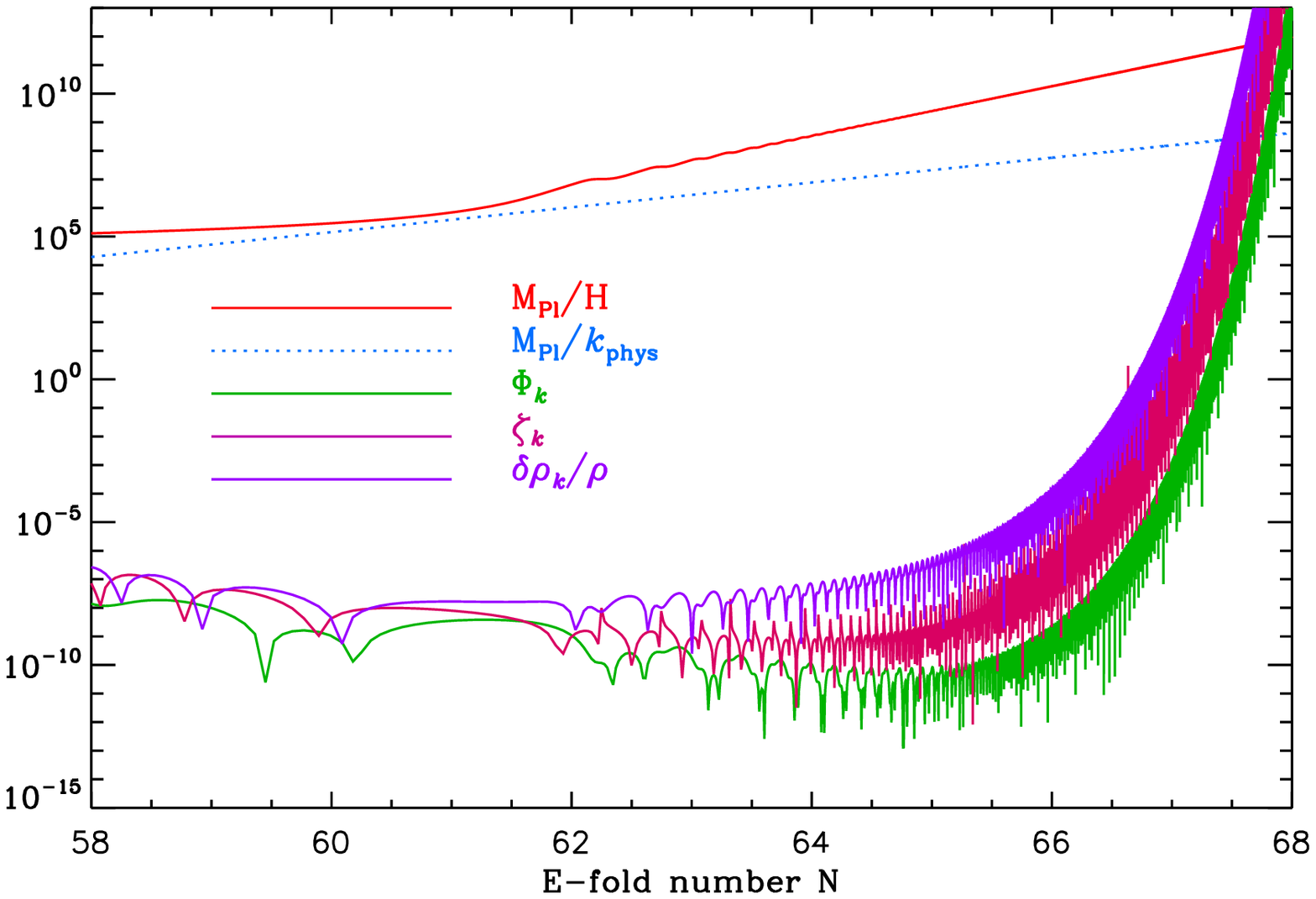}
  \end{tabular}
  \caption[...]{Evolution of the density perturbations
    $\delta\rho_{\bm k}/\rho$, $\zeta _{\bm k}$ and $\Phi_{\bm k}$
    towards the end of quartic inflation and during the ensuing period
    of metric preheating after inflation. The dimensionless parameter
    is $\lambda \simeq 10^{-12}$. The red solid line represents the
    Hubble radius while the blue dotted line represents the physical
    wavelength of the Fourier mode. Left panel: the Fourier mode is
    characterized by $\hat{k}\simeq 1.427$ and lies outside the
    resonance band. Right panel: $\hat{k}\simeq 1.738$ and the mode is
    inside the instability band as the exponential growth indicates.}
\label{fig:modequartic}
\end{figure*}

In the previous section, we have studied a large field model with a
quadratic potential. This corresponds to a pretty generic situation
since, close to its minimum, the inflaton potential can always be
approximated by a parabola. However, it might also be that, for some
unspecified reason (generically, the presence of a particular
symmetry), the quadratic term in the Taylor expansion~(\ref{eq:pot})
is absent. Then, one has to repeat the previous analysis in the case
where the potential is given by
\begin{equation}
\label{eq:potquartic}
V(\phi)=\frac{\lambda }{4}\phi ^4 \, .
\end{equation}
In the following, we find that growth also occurs for this case. This
result is particularly interesting, as it shows that the growth of
density perturbations goes beyond the simple analogue of Newtonian
collapse in a dust Universe. Indeed, in the case of $\lambda\phi^4/4$,
the effective equation of state is that of radiation.

\par

As before, we assume that the potential is approximated by the above
quartic form in the preheating era without making any specific
assumption on the actual potential during the inflationary era. We
note that quartic inflation can be normalized to COBE anisotropies
with $\lambda \simeq 10^{-12}$, but that at the same time such an
inflationary potential is disfavored by detailed analysis of CMB
data~\cite{Martin:2006rs}.

\par

For a quartic potential, the evolution of the inflaton field around
its minimum can be expressed in the large time limit $t\,\gg\,H_{\rm
  end}^{-1}$
as~\cite{Kofman:1997yn,Kaiser:1997mp,Kaiser:1997hg,Bassett:2005xm}
\begin{equation}
\phi (\eta )=\phi _{\rm end}\left(\frac{a_{\rm end}}{a}\right)\mbox{cn}
\left(x-x_{\rm end};\frac{1}{\sqrt{2}}\right)\, ,
\end{equation}
where $x\equiv a_{\rm end}\phi_{\rm end}\sqrt{\lambda}\eta $, $\eta $
being the conformal time and $\mbox{cn}$ denotes the Jacobi cosine
function~\cite{Abramovitz:1970aa,Gradshteyn:1965aa}. This solution
should be inserted in Eq.~(\ref{eq:evolution}). Then, keeping only the
dominant terms, the equation that controls the evolution of Mukhanov's
variable reduces to
\begin{eqnarray}
\ddot{\tilde{v}}_{\bm k} &+& \Biggl[\frac{k^2}{a^2}
+3\lambda \phi_{\rm end}^2\left(\frac{a_{\rm end}}{a}\right)^2
\nonumber \\ & & \times \mbox{cn}^2
\left(x-x_{\rm end};\frac{1}{\sqrt{2}}\right)\Biggr]
\tilde{v}_{\bm k}=0\, .
\label{eq:vquartiv}
\end{eqnarray}
It is more convenient to change back to the variable $v_{\bm k}$ and
to express this equation in terms of the variable $x$ defined above:
\begin{equation}
\frac{{\rm d}^2v_{\bm k}}{{\rm d}x^2}+\left[\frac{k^2}{\phi_{\rm end}^2
a_{\rm end}^2\lambda}+3\mbox{cn}^2
\left(x-x_{\rm end};\frac{1}{\sqrt{2}}\right)\right]v_{\bm k}=0\, .
\end{equation}
As emphasized in
Refs.~\cite{Kaiser:1997mp,Kaiser:1997hg,Bassett:2005xm}, this equation
is the so-called Lam\'e equation for which the Floquet chart is
exactly known. There is amplification if $3/2<k^2/(\phi_{\rm end}^2
a_{\rm end}^2\lambda)<\sqrt{3}$. Using $H_{\rm
  end}^2\simeq \kappa \lambda \phi_{\rm end}^4/12$, the above
inequality can be cast into the following form
\begin{equation}
\label{eq:band}
\frac{3}{2\sqrt{\pi}}\frac{m_{\rm Pl}}{\phi_{\rm
    end}}\,<\,\hat{k}\,<\,
\left(\frac{3\sqrt{3}}{2\pi}\right)^{1/2}\frac{m_{\rm
  Pl}}{\phi_{\rm end}}\, .
\end{equation}
For the particular case of quartic large field inflation, $\phi_{\rm
  end}/\mpl\simeq 1/\sqrt{\pi}$, hence the lower and upper bound of
the above band reduce respectively to $3/2$ and $(3\sqrt{3}/2)^{1/2}$.

\par

For the above instability, the maximum growth occurs with a Floquet
index $\mu\,\simeq\, 0.036$ at $k^2/(\phi_{\rm end}^2 a_{\rm
  end}^2\lambda)\,\simeq\, 1.615$ (\ie $\hat k\simeq 1.56$ for a large
field quartic inflationary model)~\cite{Bassett:2005xm}. The Floquet
index is constant, therefore growth becomes exponential, unlike what
has been found for the quadratic case.

\par

Several remarks are in order at this point. Firstly, the resonance
band corresponds to a fixed narrow range of comoving scales, contrary
to the quadratic case for which the band widens with time. Secondly,
the modes that are subject to parametric resonance are of spatial
scale comparable with the Hubble radius at the end of inflation.
Thirdly, the effect found here is similar to that investigated in
Ref.~\cite{Kofman:1997yn,Bassett:2005xm} except that it concerns
Mukhanov's variable rather than the perturbed inflaton field or the
Fourier amplitude of a scalar field $\chi $ coupled to the inflaton as
$g^2\phi^2\chi^2/2$, where $g$ is the coupling constant. The case of
density perturbations can be viewed as a special case with a specific
value of $g$. But, of course, the underlying physics is different in
the sense that metric fluctuations are different from an external
scalar field coupled to $\phi$.

\par

In order to check the effect discussed above, we have numerically
integrated the equation of motion for a large field inflationary
quartic potential, see Fig.~\ref{fig:modequartic}. The numerical
analysis confirms the previous analytical analysis, although we find
that the instability band is shifted by $12\,$\% to higher values in
comparison with Eq.~(\ref{eq:band}); this effect is likely due to our
assumption of instantaneous transition from inflation to a (radiation
dominated) preheating era in the analytical calculations. In detail,
one finds the band at $1.684<\hat{k}<1.807$. When the mode is outside
the resonance band (left panel in Fig.~\ref{fig:modequartic}), the
perturbations remain constant while when it is inside (right panel in
Fig.~\ref{fig:modequartic}), we observe a violent amplification of
$\Phi _{\bm k}$, $\zeta_{\bm k}$ and $\delta _{\bm k}$.

\par

To summarize, the quartic case presents substantial differences in
comparison with the quadratic case. There exists a narrow instability
band at scales of order of the Hubble radius at the end of inflation,
of fixed width (in terms of comoving wavenumber) and with associated
exponential growth. 

\subsection{The mixed quadratic-quartic case}
\label{sec:mixedcase}

\begin{figure*}
  \centering
  \begin{tabular}{cc}
\includegraphics[width=0.45\textwidth,clip=true]{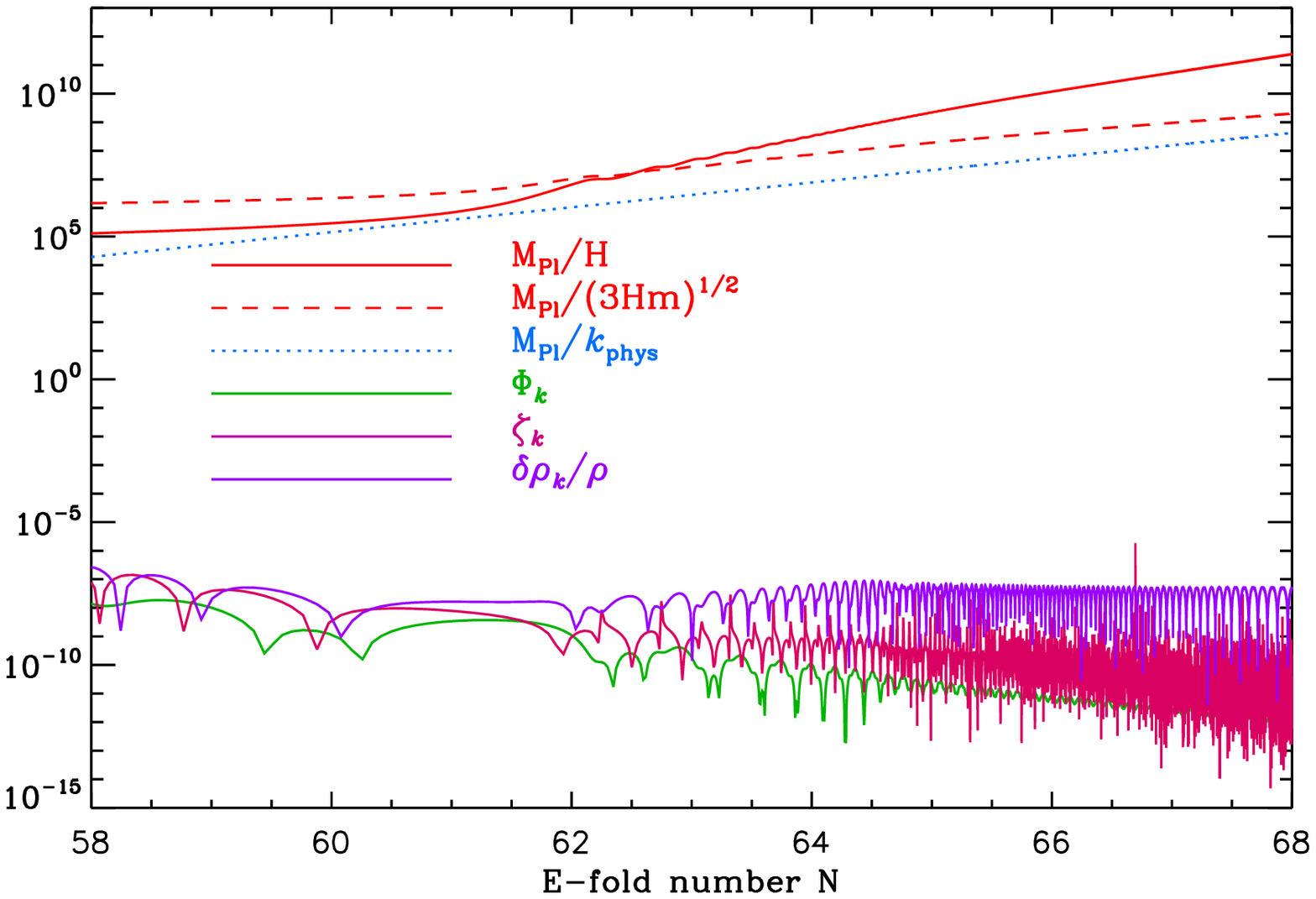} 
\includegraphics[width=0.45\textwidth,clip=true]{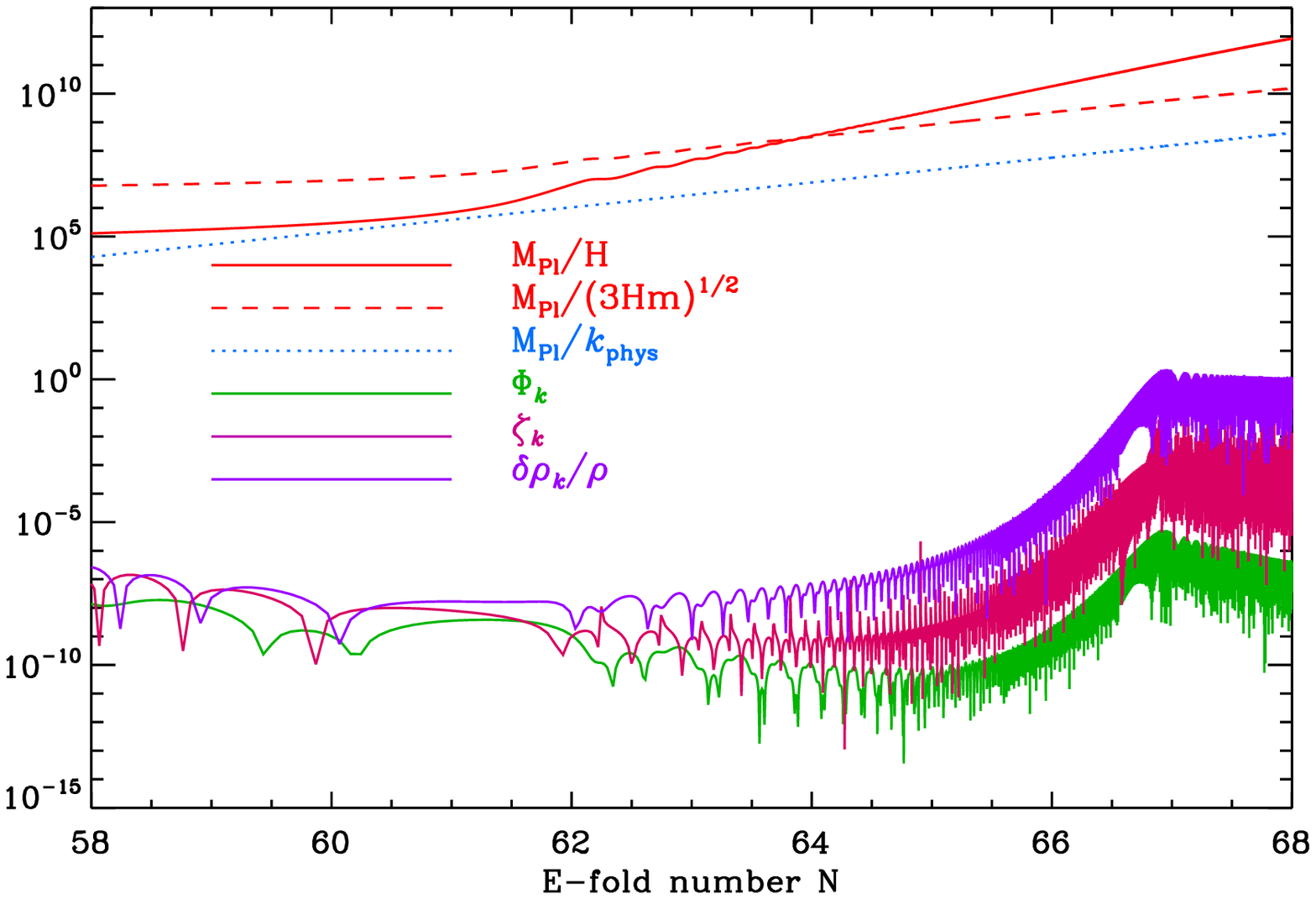} 
    \end{tabular}
    \caption[...]{Left panel: evolution of the density perturbations
      $\delta\rho_{\bm k}/\rho$, $\zeta _{\bm k}$ and $\Phi_{\bm k}$
      towards the end of inflation with a potential given by
      Eq.~(\ref{eq:potmixed}) and during the ensuing period of of
      metric preheating. The dimensionless parameter are $\lambda
      \simeq 10^{-12}$ and $\Upsilon =0.05$. The red solid line
      represents the Hubble radius, the red dashed line the scale
      $\ell _{_{\rm C}}$ while the blue dotted line represents the
      physical wavelength of the Fourier mode. The Fourier mode is
      characterized by $\hat{k}\simeq 1.738$. Right panel: same as
      left panel but with $\Upsilon=0.003$ and $\hat{k}=1.737$. The
      value of $k$ is the same for the two panels but the value of
      $\hat{k}$ sightly differs because the end of inflation does not
      occur exactly at the same time in the two cases.}
\label{fig:modemixed}
\end{figure*}

The purely quartic case arises only if the quadratic term is absent in
the potential. Here we briefly address the more general case, in which
both quadratic and quartic terms are present:
\begin{equation}
\label{eq:potmixed}
V(\phi)=\frac{1}{2}m^2\phi^2+\frac{\lambda }{4}\phi^4\, .
\end{equation}
At large vacuum expectation values of the inflaton, the quartic term
always dominates while at small values of $\phi$, the quadratic term
is the most important one. For which value of $\phi $ these terms are
of equal importance depends on the two free parameters $m$ and
$\lambda$. If the potential were purely quartic, then inflation would
stops at $\phi_{\rm end}=\mpl/\sqrt{\pi}$. For this value, the two
terms are equal if $m=\sqrt{\lambda /(2\pi )}\mpl $. It is therefore
convenient to parameterize the potential~(\ref{eq:potmixed}) by
$\lambda $ and the dimensionless parameter $\Upsilon$ defined by
\begin{equation}
\Upsilon\equiv \frac{m}{\mpl}\sqrt{\frac{2\pi }{\lambda }}\, .
\end{equation} 
The regime where $\Upsilon \gg 1$ reduces to the quadratic case while
the case $\Upsilon \ll 1$ is equivalent to the quartic case. The most
interesting situation is of course when $\Upsilon $ is of order one.

\par

In order to study this case, we have numerically integrated the
equations of motion, assuming that the potential form remains the same
during inflation and the preheating era. As a first example, we have
taken $\lambda =10^{-12}$, that is to say the same value as in the
previous section for numerical applications and $\Upsilon=0.05$. We
have also chosen $\hat{k}=1.738$, \ie a value which is inside the
resonance band in the pure quartic case, see right panel in
Fig.~\ref{fig:modequartic}. The result is displayed in
Fig.~\ref{fig:modemixed} (left panel). Several comments are in
order. Firstly, one notices that the Hubble radius and the
characteristic scale $\ell _{_{\rm C}}$ no longer coincide. This is of
course due to the fact that the mass $m$ is no longer COBE
normalized. In fact, in this model, the end of inflation is delayed
because at $\phi=\mpl/\sqrt{\pi}$, the quadratic term still plays a
role. Regarding the perturbations themselves, despite the fact that
the Fourier mode is inside the quartic resonance band of
Eq.~(\ref{eq:band}), there is no amplification. It is of course due to
the fact that, for small vacuum expectation values of the field, the
dominant term is the quadratic one. We also notice that $\delta
_\mathbf{k}$ does not grow as $a$ either. This is because, since the
end of inflation has been postponed and the value of $\ell_{_{\rm C}}$
increased (in comparison with the corresponding value in the pure
quadratic case), the time at which the mode crosses $\ell _{_{\rm C}}$
is also delayed (and therefore lies outside of the plot).

\par

It is clear that if we decrease the value of $\Upsilon$, and keep the
same value for $k$, we should observe the growth described in the last
section in the pure quartic case. This is indeed the case as
demonstrated in Fig.~\ref{fig:modemixed} (right panel) where
$\Upsilon=0.003$. Just after the end of inflation, the Fourier mode is
inside the pure quartic resonance band and the perturbations grow. But
as the amplitude of the inflaton field decreases, the influence of the
quadratic term becomes more and more important. As a consequence, the
exponential growth stops at some point. Then, the amplitude of the
fluctuations remain constant until the mode exceeds the characteristic
scale $\ell _{_{\rm C}}$ associated to the quadratic part of the
potential. As before, this time is delayed due to the small value of
the mass $m$ (small in comparison with the COBE normalized value) and
lies outside of the plot. Then, the density contrast grows
proportional to the scale factor as discussed in
Section~\ref{sec:mcarrephicarre}.

\section{Conclusion}
\label{sec:conclusions}

In this paper we have investigated the evolution of sub-horizon
perturbations between the end of cosmic inflation and the beginning of
a radiation dominated early Universe. This study assumes that energy
density of the Universe during this phase is dominated by the
oscillations of a single scalar field around its minimum in a
quadratic or quartic potential.

\par

We have found that perturbations are subject to a preheating
instability in the narrow resonance regime, leading to the growth of
the density contrast $\delta\rho_{\bm k} /\rho$ on sub-horizon scales.
In the case of a quadratic potential, the domain of instability (in
terms of comoving wavenumbers) widens with time. This implies that a
wavenumber that has entered the band at some time remains inside for
the whole duration of the preheating era. The density contrast of
sub-horizon unstable modes then increases as the scale factor,
reaching non-linearity for a wide range of scales. Maximum growth
occurs for spatial scales of the order of the Hubble radius at the end
of inflation, to which is associated a mass scale $M_{\rm end}\simeq
3.3\times 10^3\,\mbox{g}\,\left(H_{\rm end}/10^{13}\,{\rm
    GeV}\right)^{-1}$.

\par

A quartic potential leads to a different phenomenology, although the
main effect of preheating instability remains. Notably, the range of
unstable wavenumbers is now fixed in terms of comoving wavenumbers and
peaked around the Hubble scale at the end of inflation. On the other
hand, the density contrast of sub-horizon unstable modes increases
exponentially fast.

\par

The growth of structure on sub-horizon scales is known to occur in a
matter dominated era, hence the growth of the density contrast for a
quadratic inflaton potential can thus be well understood in this
respect. However, the result obtained for a quartic potential confirms
that the effect goes beyond a simple analogy with Newtonian collapse
of a self-gravitating dust cloud, as the effective equation of state
in this latter case is that of radiation.

\par

These results open interesting avenues of research in the search for
observational signatures of the dynamics of the very early
Universe. In particular, as we have noted, the abundance of black
holes of mass comparable to the above is constrained by cosmological
arguments. The formation of inflaton halos may also lead to copious
production of gravitational waves, as discussed in a companion
paper. Finally, the very existence of non-linearities on small scales
casts some shadow on the reliability of the predictions for curvature
perturbations on those large scales responsible for the present
large-scale structure. These latter could be modified if effects of
mode-mode coupling become important.

\bibliographystyle{h-physrev2}
\bibliography{biblio_preheat}
\end{document}